\documentclass[twocolumn]{aastex62}
\usepackage{xcolor}
\usepackage{apjfonts}

\newcommand{\caat}{$Ca_{\rm AT}$}

\newcommand{\cacnch}{Ca--CN--CH}
\newcommand{\cactio}{$Ca_{\rm CTIO}$}
\newcommand{\caii}{\ion{Ca}{2}}
\newcommand{\cajwl}{$Ca_{\rm JWL}$}
\newcommand{\chjwl}{$ch_{\rm JWL}$}
\newcommand{\cnjwl}{$cn_{\rm JWL}$}
\newcommand{\cnjwlcor}{$cn_{\rm JWL,cor}$}
\newcommand{\cnpjwl}{$cn^\prime_{\rm JWL}$}
\newcommand{\cnw}{CN-w}
\newcommand{\cns}{CN-s}

\newcommand{\hkpjwl}{$hk^{\prime}_{\rm JWL}$}
\newcommand{\hst}{{\it HST}}
\newcommand{\nrgb}{$n$(\cnw):$n$(\cns)}

\newcommand{\pdca}{$\parallel$$\Delta Ca$}
\newcommand{\pchjwl}{$\parallel$$ch_{\rm JWL}$}
\newcommand{\pcnjwl}{$\parallel$$cn_{\rm JWL}$}
\newcommand{\pcnpjwl}{$\parallel$$cn^\prime_{\rm JWL}$}
\newcommand{\phkctio}{$\parallel$$hk_{\rm CTIO}$}
\newcommand{\phkjwl}{$\parallel$$hk_{\rm JWL}$}
\newcommand{\phkpjwl}{$\parallel$$hk^{\prime}_{\rm JWL}$}
\newcommand{\str}{Str{\"o}mgren}
\newcommand{\vbump}{$V_{\rm bump}$}

\newcommand{\vhb}{$V_{\rm HB}$}
\newcommand{\vvhb}{$V - V_{\rm HB}$}
\newcommand{\vvhbmag}{$-$2 mag $\leq$ $V - V_{\rm HB}$ $\leq$ 2 mag}
\newcommand{\cnwave}{$\lambda$3883}
\newcommand{\chwave}{$\lambda$4250}

\newcommand{\ds}{$\delta S$(3839)}

\newcommand{\sch}{CH(4300)}

\newcommand{\cubi}{$C_{UBI}$}

\newcommand{\hkjwl}{$hk_{\rm JWL}$}
\newcommand{\hkctio}{$hk_{\rm CTIO}$}
\newcommand{\gaia}{{\it Gaia}}

\newcommand{\ciso}{$^{12}$C/$^{13}$C}
\newcommand{\agbm}{AGB-manqu{\'e}}

\shorttitle{\cnjwl\ versus \cnpjwl}
\shortauthors{Lee}
\begin{document}

\title{Multiple Stellar Populations of Globular Clusters from Homogeneous \cacnch\ Photometry. V. \cnpjwl\ as a Surrogate \cnjwl\ Index and NGC\,6723.\footnote{Based on observations made with the Cerro Tololo Inter-American Observatory (CTIO) 1 m telescope, which is operated by the SMARTS consortium, and the Kitt Peak National Observatory (KPNO) 0.9 m telescope, which is operated by WIYN Inc. on behalf of a Consortium of partner Universities and Organizations.}
\footnote{This work has made use of data from the European Space Agency (ESA) mission \gaia\ (\url{https://www.cosmos.esa.int/gaia}), processed by the \gaia\ Data Processing and Analysis Consortium (DPAC, \url{https://www.cosmos.esa.int/web/gaia/dpac/consortium}). Funding for the DPAC has been provided by national institutions, in particular the institutions participating in the \gaia\ Multilateral Agreement.}
}

\author[0000-0002-2122-3030]{Jae-Woo Lee}
\affiliation{Department of Physics and Astronomy, Sejong University\\
209 Neungdong-ro, Gwangjin-Gu, Seoul, 05006, Korea\\
jaewoolee@sejong.ac.kr, jaewoolee@sejong.edu}

\begin{abstract}
We introduce new color indices \cnpjwl\ (= $Ca_{\rm CTIO} - Ca_{\rm JWL}$) and \chjwl\ [=$(JWL43 - b) - (b-y)$], accurate photometric measures of the CN band at \cnwave\ and the CH G band, respectively, in the study of the multiple populations (MPs) in globular clusters (GCs). Our photometric CN--CH relation for a large number of red-giant branch (RGB) in M5 shows that the evolutions of the CN and CH between the \cnw\ and \cns\ populations are not continuous. Armed with our new color indices, we investigate the MPs of NGC\,6723, finding the RGB populational number ratio of \nrgb\ $\approx$ 35.5:64.5 ($\pm$2.8) with no radial gradient. Similar to other normal GCs with MPs, the helium abundance of the \cns\ population inferred from the RGB bump magnitude is slightly enhanced by $\Delta Y$ = 0.012 $\pm$ 0.012. Our \cnjwl\ and \cnpjwl\ color-magnitude diagrams clearly show the discrete double AGB populations in NGC\,6723, whose bright AGB populational number ratio is in marginally agreement with that of the RGB stars within the statistical errors.
Finally, our synthetic \cnjwl\ index is in good agreement with observations, except for the \cnw\ asymptotic giant branch (AGB). To mitigate the discrepancy in the \cnw\ AGB may require a mild nitrogen enhancement and/or a large decrement in the \ciso\ ratio with respect to the bright RGB.
\end{abstract}


\section{Introduction}
The discovery that the normal globular clusters (GCs) in our Galaxy contain multiple populations (MPs) demanded a paradigm shift and would surely be one of the most greatest achievements in stellar and galactic astrophysics during the past decade \citep[e.g., see][]{carretta09,jwlnat,piotto15}.
In observational perspective, one of the important issues is an increasing need for the clear and simple definition of MPs in GCs. In low-resolution spectroscopy, the well-known ubiquitous nature of the CN bimodal distributions in GCs can serve a pivotal role in populational tagging, since the discrete multiple sequences resulted from the nitrogen enhancement of the later generation of the stars is a hallmark of MPs in GCs.
In high-resolution stellar spectroscopy, the assignment of MPs had been usually performed on the Na--O plane, where the evolutions of oxygen and sodium between MPs are appear to be continuous mainly due to the large uncertainties in spectroscopic measurements \citep[e.g., see][]{marino08,lee10,lee16,carretta14}.
Also, the spectroscopic study of MPs had been always restricted to bright stars and can not be applied to crowded fields, such as the central part of GCs in which most of GC membership stars reside.

The populational tagging in stellar photometry can be more complex and subtle. In our previous works \citep{lee17,lee18,lee19a}, we assessed the utility of various photometric systems used in the study of MPs in GCs, finding that some of the widely used photometric systems, such as the \str\ $m1$, $cy$ indices, the Sloan Digital Sky Survey (SDSS) photometric system, and the \cubi\ [=$(U-B)-(B-I)$] index, for example, may not be well-suited for the accurate populational tagging of individual stars in GCs with MPs.

In this paper, we proceed further and introduce a new set of the photometric index that can be very useful in the study of MPs in GCs: \cnpjwl\ (= $Ca_{\rm CTIO} - Ca_{\rm JWL}$) and \chjwl\ [=$(JWL43 - b) - (b-y)$]. They are accurate photometric measures of the CN band at \cnwave\ and the CH G band at \chwave, respectively. As we will show later, the \cnpjwl\ index is an excellent surrogate \cnjwl.

Armed with new set of photometric indices, we investigate the MPs of NGC\,6723. \citet{lim16} reported an interesting result that the cumulative radial distribution of the \cnw\ red-giant branch (RGB) population of the cluster is more centrally concentrated than that of the \cns\ population is, leading them to argue that an observational line of the counterexample to the currently accepted formation scenario for GCs with MPs \citep[e.g., see][]{bekki10}.
We will show later that both the \cnw\ and \cns\ populations in NGC\,6723 have identical cumulative radial distributions and the results presented by \citet{lim16} were in error, mainly due to their incorrect populational assignment of the individual stars from a poor photometric CN tracer.

The outline of this paper is as follows.
In Section \ref{s:2}, we show the definitions of photometric indices used in our study.
In Section \ref{s:3}, we discuss our new photometric index, \chjwl, and the CH band contamination of the continuum sidebands of the spectroscopic measurements of the CH G band will be discussed. 
In Section \ref{s:4}, the utilities of our photometric indices, such as a photometric CN--CH anticorrelation in M5 RGB stars is discussed.
In Section \ref{s:5}, we discuss the CH band contamination in the passband of the \ion{Ca}{2} H \& K region.
Finally, in Section \ref{s:6}, we discuss some aspects of the MPs of the inner halo GC NGC\,6723, including the He enhancement in the \cns\ population, identical cumulative radial distributions in individual populations and a hint for the nitrogen enhancement and the decrement in the carbon isotopes ratio in the \cnw\ asymptotic giant branch (AGB), is discussed.

\section{Photometric Indices for Multiple Stellar Populations}\label{s:2}
\begin{figure}
\epsscale{1.2}
\figurenum{1}
\plotone{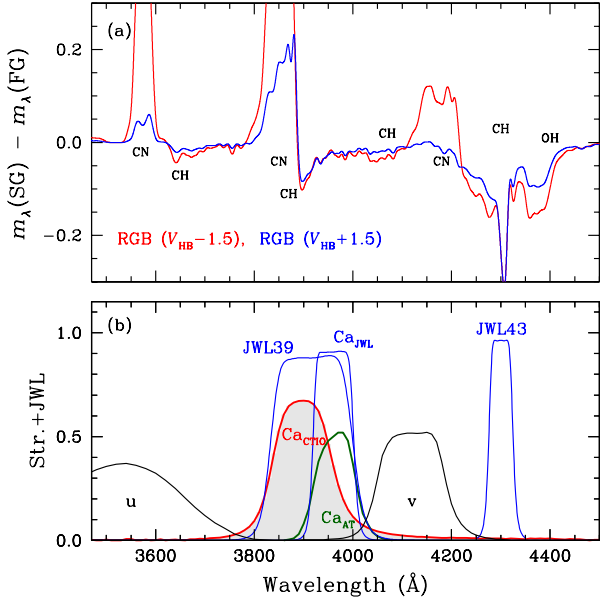}
\caption{
(a) Comparisons of synthetic spectra for RGB stars with intermediate metallicity. The red solid line shows the difference in the monochromatic magnitude between the bright [$V = V_{\rm HB} - 1.5$ mag, where $V_{\rm HB}$ is the $V$ magnitude of the horizontal branch (HB)] \cns\ and \cnw\ RGB stars, $\Delta m_\lambda$ =  $m_\lambda$(\cns) $-$ $m_\lambda$(\cnw), and the blue solid line shows that for the faint RGB stars ($V = V_{\rm HB} + 1.5$ mag).
(b) The filter transmission curves of the extended \str\ and the JWL systems.
The \str\ $u$ and $v$ filters are shown with black solid lines, \cajwl, $JWL39$ and $JWL43$ are shown with blue solid lines.
The calcium filter provided by the CTIO, \cactio, is denoted by the red solid line with shaded-area and the calcium filter defined by \citet{att91}, $Ca_{\rm AT}$, is denoted by the dark green line.
}\label{fig:filter}
\end{figure}

In Figure~\ref{fig:filter}, we show the transmission curves of the \str\ and the JWL systems along with the $Ca$ filter provided by the CTIO, \cactio, and the $Ca$ filter by \citet{att91}, \caat.
The \cajwl, $JWL39$ and $JWL43$ filters were developed by the author of the paper in order to study the MPs in GCs \citep[e.g., see][]{lee15,lee17,lee18,lee19a}.
As we already noted in our previous work, the \cajwl\ filter was designed to have a filter bandwidth and pivot wavelength very similar to that of the F395N filter in the Wide-Field Camera 3 (WFC3) on board the {\it Hubble Space Telescope} (\hst) and it can measure the \caii\ H and K absorption strengths.
On the other hand, our $JWL39$ filter was designed to measure the CN band at \cnwave\ in combination with our \cajwl.
Finally, our $JWL43$ filter was designed to measure the CH band at \chwave\ in combination with \str\ $b$ and $y$, as will be described below.

The original \cactio\ was designed to have a filter bandwidth and pivot wavelength very similar to that of the \caat\ defined by \citet{att91}, but it had been degraded due to aging, and its original transmission curve had been significantly altered to the shorter wavelength, resulting in having the passband similar to our $JWL39$ filter. But it should be pointed out that our $JWL39$ filter has a more uniform and high transmission across the passband, dropping more rapidly at both edges. As a consequence, the measurements made through the \cajwl\ and \cactio\ filters can be slightly different.

In our work here, we will use \emph{seemingly similar but different} sets of photometric index, some of which are already defined elsewhere \citep[e.g., see][]{att91,lee15,lee17},
\begin{eqnarray}
hk &=& (Ca_{\rm AT}-b) - (b-y), \label{eq:hk} \\
hk_{\rm CTIO} &=& (Ca_{\rm CTIO}-b) - (b-y), \label{eq:hkctio} \\
hk_{\rm JWL} &=& (Ca_{\rm JWL}-b) - (b-y), \label{eq:hkjwl} \\
hk^\prime_{\rm JWL} &=& (JWL39 - b) - (b-y),  \label{eq:hkpjwl} \\
cn_{\rm JWL} &=& JWL39 - Ca_{\rm JWL}, \label{eq:cn} \\
cn^\prime_{\rm JWL} &=& Ca_{\rm CTIO} - Ca_{\rm JWL},  \label{eq:cnp} \\
ch_{\rm JWL} &=& (JWL43 - b) - (b-y). \label{eq:ch}
\end{eqnarray}
The $hk$ and the \hkjwl\ indices are known to be a good photometric measure of calcium abundances at a given luminosity or color \citep{att91,jwlnat,lee09,lee15}.
Later, it will be shown that the filter bandwidths of the \caat\ and the \cajwl\ contain weak CH molecular absorption lines and, as a consequence, they are likely affected by weak CH band contamination.

Similar to our previous work \citep{lee19a}, the (double) RGB sequences in the individual color indices were parallelized using the following relation,
\begin{equation}
\parallel{\rm CI}(x) \equiv \frac{{\rm CI}(x) - {\rm CI}_{\rm red}}
{{\rm CI}_{\rm red}-{\rm CI}_{\rm blue}},\label{eq1}\label{eq:pl}
\end{equation}
where, CI$(x)$ is the color index of the individual stars and CI$_{\rm red}$, CI$_{\rm blue}$ are color indices for the fiducials of the red and the blue sequences.
Note that the `$\parallel$' operator is the same as the `$\Delta_r$' operator that we already defined in our previous work \citep{lee19a}.

\begin{figure}
\epsscale{1.2}
\figurenum{2}
\plotone{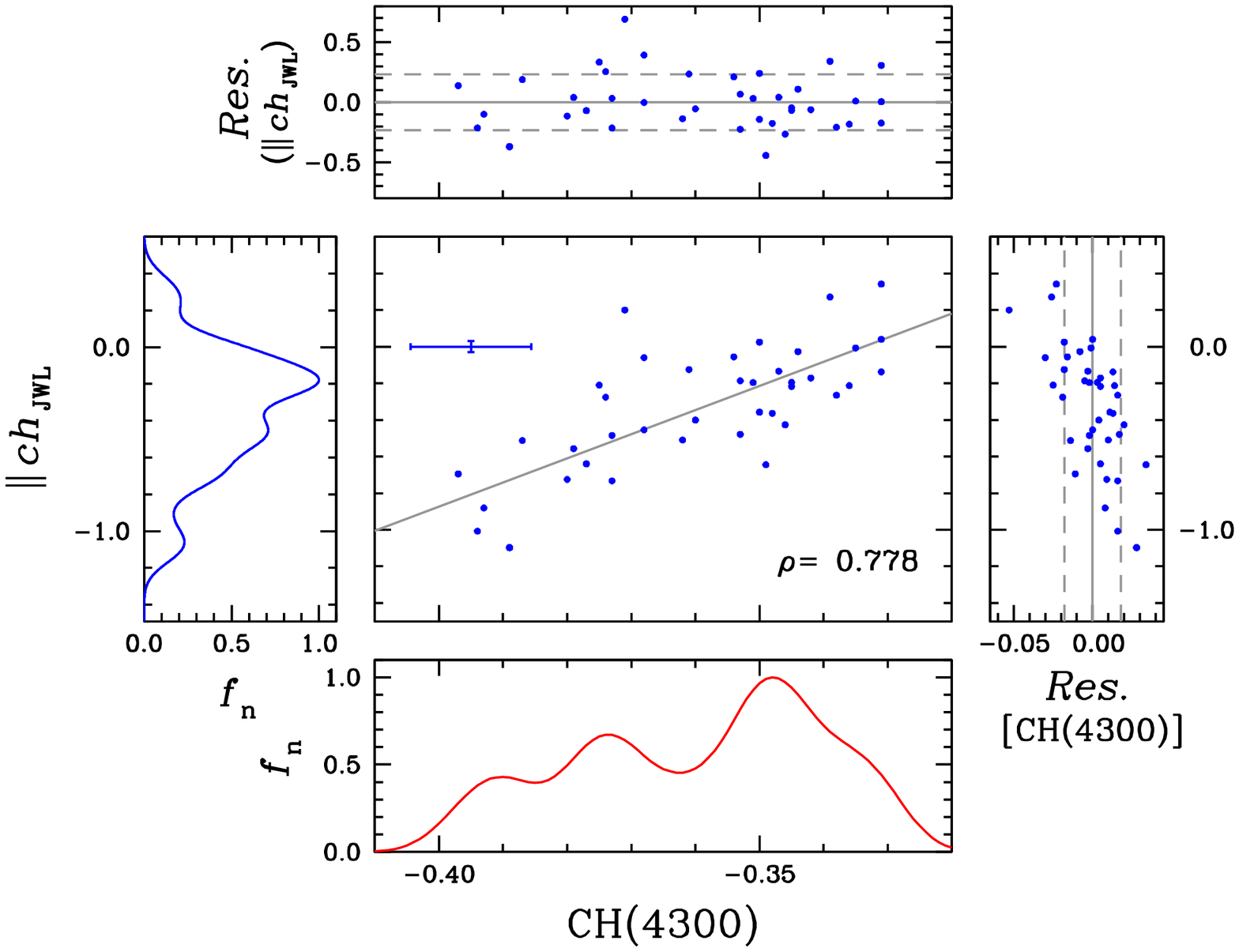}
\caption{
Plot of \sch\ versus \pchjwl\ for M3 RGB stars along with the least-square fit.
The mean residuals ($\pm 1\sigma$) around the fitted lines are also shown with long-dashed lines.
The \pchjwl\ index is decently correlated with the spectroscopic \sch\ index measured by \citet{smolinski11}.
The scatter around the fitted line is $\sigma$ = 0.232 and  it is at the level of 7.7$\times\sigma$(\pchjwl), where $\sigma$(\pchjwl) is the photometric measurement uncertainty.
In the right panel, we show the residuals in \sch\ around the fitted line, finding $\sigma$ = 0.018, and it is at the level 1.9$\times\sigma$[\sch], where $\sigma$[\sch] is the spectroscopic measurement uncertainty reported by \citet{smolinski11}.
}\label{fig:m3ch}
\end{figure}

\section{\chjwl\ as a Photometric Measure of the CH G Band}\label{s:3}
As shown in Figure~\ref{fig:filter}, our $JWL43$ filter is carefully designed to measure the CH G band strength at \chwave.
In our previous works, we elaborately showed that our \cnjwl\ index is truly a measure of the CN band strengths at \cnwave\ with great accuracy.
Here, we assess the \chjwl\ index [= $(JWL43-b)-(b-y)$] as a photometric measure of the CH band strength at \chwave.

In Figure~\ref{fig:m3ch}, we show a comparison of our photometric \pchjwl\ index against the spectroscopic \sch\ index of M3 RGB stars measured by \citet{smolinski11}.
It can be said that our \pchjwl\ index is decently correlated with the \sch\ index, with the correlation coefficient of $\rho$ = 0.773.
The scatter around the fitted line is $\sigma$ = 0.232 and it is at the level of 7.7$\times\sigma$(\pchjwl), where $\sigma$(\pchjwl) is the photometric measurement uncertainty.
In the right panel, we show the residuals in \sch\ around the fitted line, finding $\sigma$ = 0.018, and it is at the level 1.9$\times\sigma$[\sch], where $\sigma$[\sch] is the spectroscopic measurement uncertainty reported by \citet{smolinski11}.

\begin{figure}
\epsscale{1.2}
\figurenum{3}
\plotone{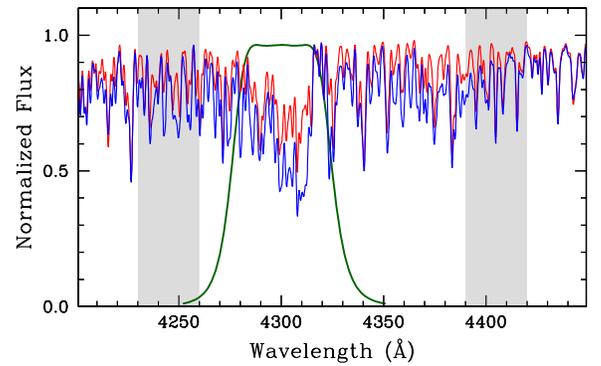}
\caption{
Comparisons of synthetic spectra of the FG (the blue solid line) and the SG (the red solid line) around the CH G band.
The continuum sidebands defined by \citet{sglee99} are shown with gray shaded boxes, showing that these continuum sidebands are severely contaminated by the CH molecular band.
The transmission function of our $JWL43$ filter is also shown with the dark green solid line.
}\label{fig:gband}
\end{figure}

The correlation between the \pchjwl\ and the \sch\ indices is not as good as that between the \pcnjwl\ and the \ds\ \citep[see also][]{lee19a}. 
As \citet{smolinski11} noted, there are several continuum sidebands are currently being used, owing to the difficulties in the continuum assignment for the spectroscopic \sch\ index:
The region around the CH G band does not appear to have clean continuum sidebands and the difficulty in the continuum placement could be a problem with great importance in deriving spectroscopic \sch\ index.
For example, \citet{smolinski11} adopted the continua defined by \citet{sglee99}, whose continuum sidebands appear to suffer from CH contamination as shown in Figure~\ref{fig:gband}.
In sharp contrast, our photometric \chjwl\ index is not affected by such a problem. 
As defined in Eq.\,(\ref{eq:ch}), our \chjwl\ index measures the CH G band strength and the slopes of the spectral energy distributions in rather clean regions, the \str\ $b$ and $y$ passbands \citep[see, e.g., Figure~1 of][]{lee19a}. 
Therefore, it is strongly believed that a rather poor correlation between our \pchjwl\ and the spectroscopic \sch\ is mainly due to adoption of the contaminated continuum sidebands by \citet{smolinski11}, which suffer CH contamination as mentioned above.

\begin{figure}
\epsscale{1.2}
\figurenum{4}
\plotone{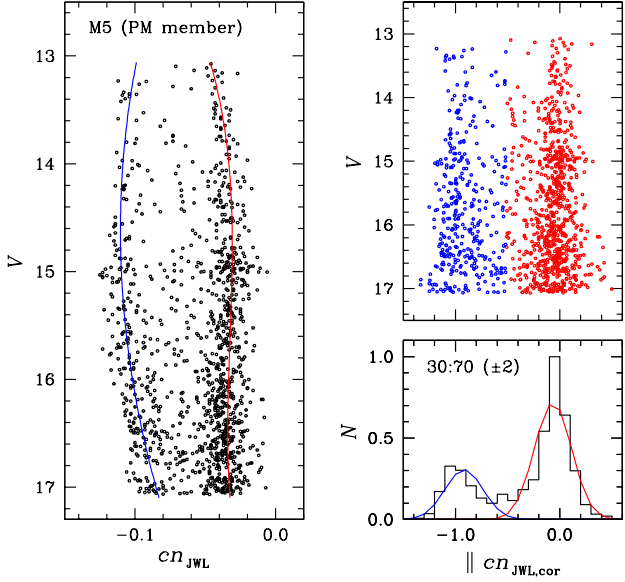}
\caption{(Left) A \cnjwl\ CMD of the proper motion membership M5 RGB stars with \vvhbmag. 
(Right) A \cnjwlcor\ CMD and the \cnjwlcor\ distribution of RGB stars. The blue color is for the \cnw\ population and the red color is for the \cns\ population.
}\label{fig:m5}
\end{figure}

\section{Color--Magnitude Diagrams for M5: Behavior of Individual Color Indices and a Photometric CN--CH Anticorrelation}\label{s:4}
In Figure~\ref{fig:m5}, we show the \cnjwl\ CMD of the M5 RGB stars with \vvhbmag.\footnote{Note that we do not have \cactio\ photometry for M3, which is not accessible at CTIO.} 
Unlike our previous studies of the cluster \citep{lee17,lee19a}, we made use of the proper motion study from the second \gaia\ date release to select the cluster's membership stars \citep{gaiadr2}, following the method similar to those used in previous works by others \citep[see, e.g.,][]{bastian18,marino18,milone18}.

Using the proper motion membership RGB stars, we derived the populational number ratio using the expectation maximization (EM) method with the two-component Gaussian mixture model.
In an iterative manner, we calculated the probability of individual stars for being the \cnw\ and the \cns\ populations, where the \cnw\ and the \cns\ populations are defined to be stars with smaller and larger \pcnjwl\ values at a given visual magnitude.
Stars with $P$(\cnw$|x_i) \geq$ 0.5 from the EM estimator correspond to the \cnw\ population, where $x_i$ denotes the individual RGB stars, while those with $P$(\cns$|x_i)$ $>$ 0.5  correspond to the \cns\ population.
Despite the incompleteness of the second \gaia\ data release in the central part of the cluster, we obtained the populational number ratio of \nrgb\ = 30:70 ($\pm$2), which is consistent with our previous results \citep{lee17,lee19a}, which is natural to expect that M5 does not show any radial gradient in the populational number ratios.

\begin{figure}
\epsscale{1.1}
\figurenum{5}
\plotone{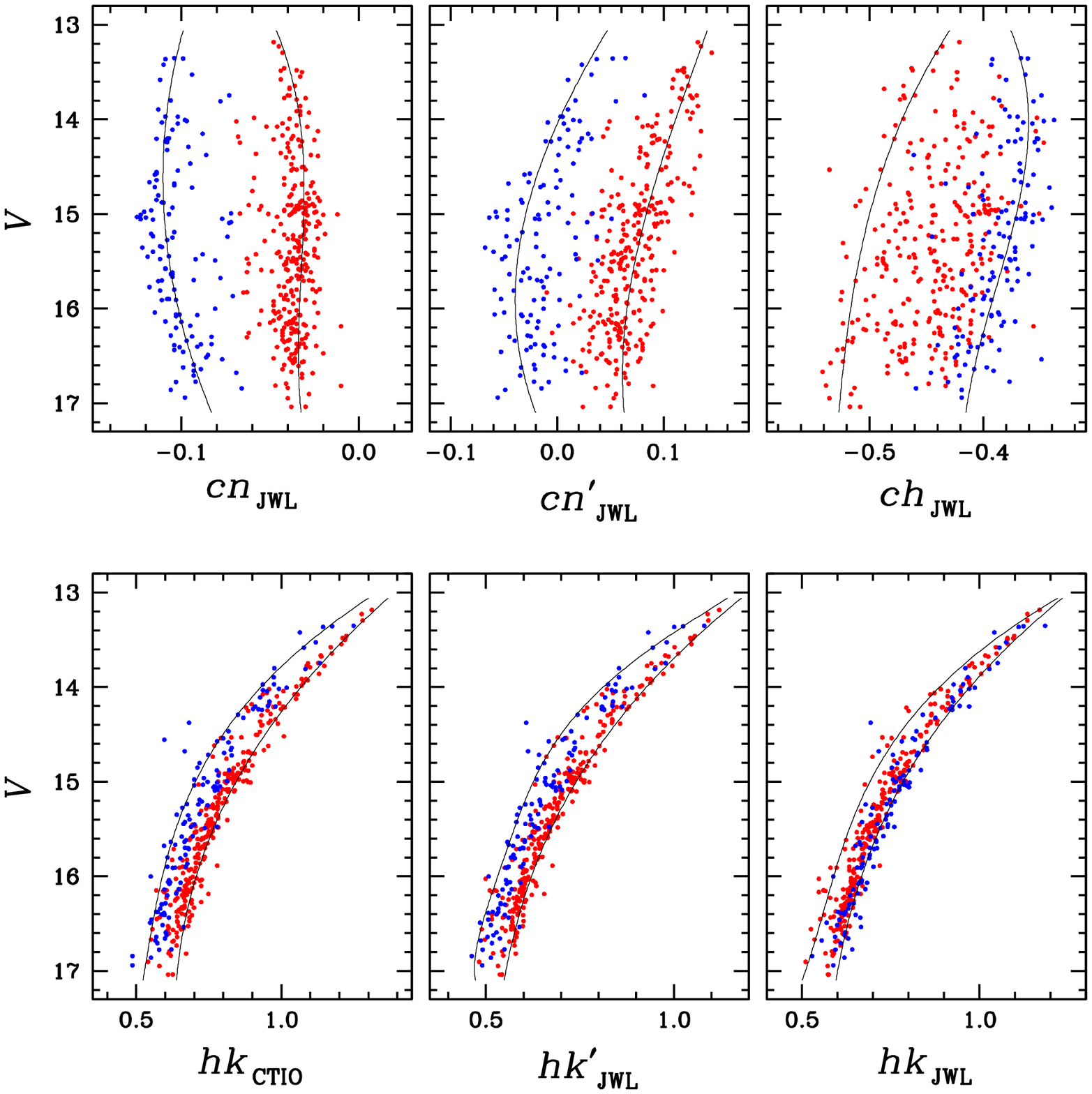}
\caption{CMDs for the proper motion membership M5 RGB stars with \vvhbmag\ and $\sigma$(\cnjwl) $\leq$ 0.005 mag.
The blue dots denote the \cnw\ population and the red dots the \cns\ population based on the \cnjwlcor\ distribution in Figure~\ref{fig:m5}. 
The black solid lines denote the fiducial sequences of individual populations in each color index.
Note that, in the \hkjwl\ CMD, the \cnw\ RGB stars are preferentially located on the red part of the \hkjwl\ (i.e.\ larger values of the \hkjwl\ index at a given $V$ magnitude), which is thought to be the contribution of the CH band absorption lines in the \cajwl\ passband.
}\label{fig:cmdidx}
\end{figure}

\begin{figure}
\epsscale{1.2}
\figurenum{6}
\plotone{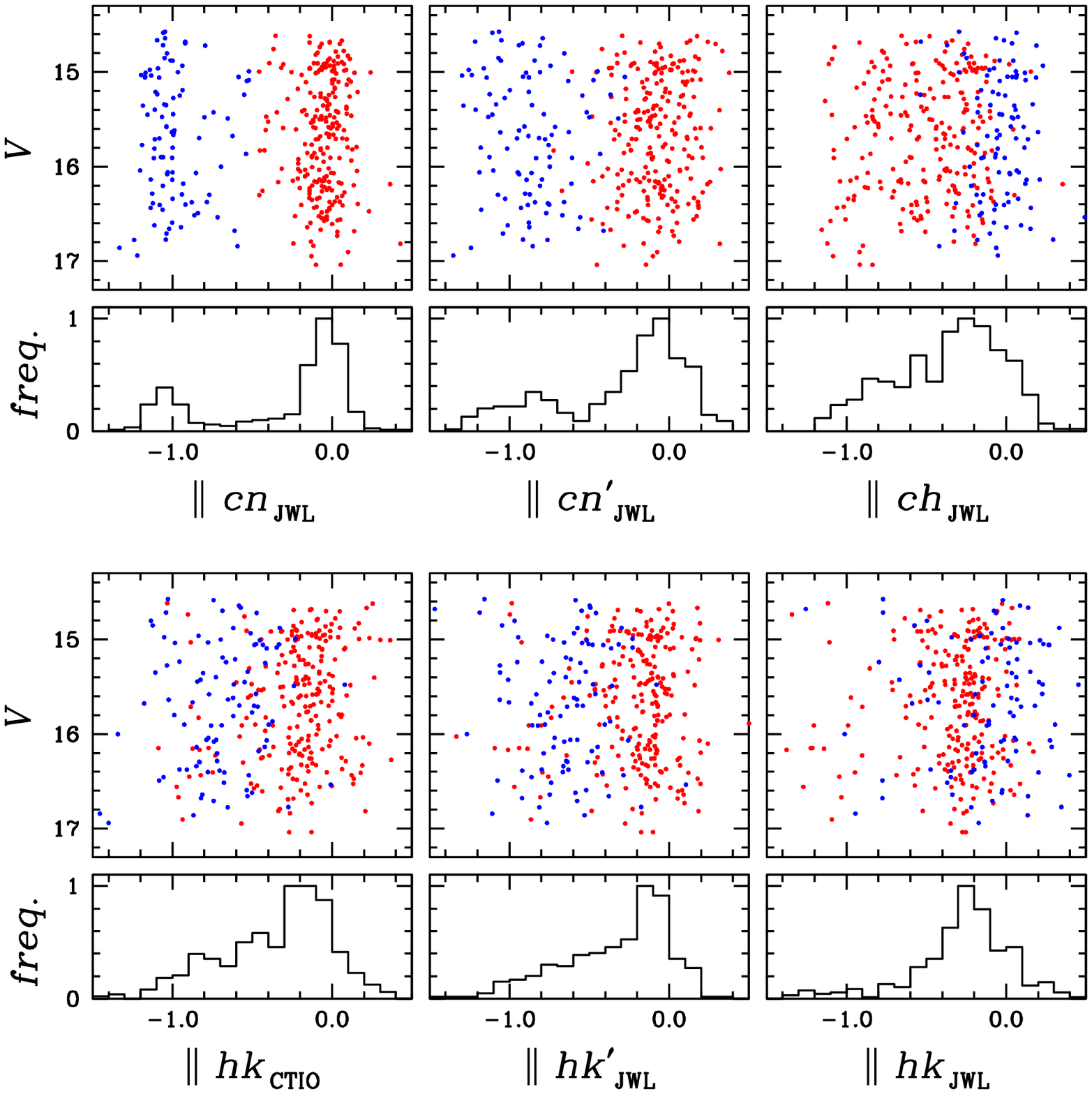}
\caption{
Parallelized CMDs for the proper motion membership M5 RGB stars with $-$0.5 $\leq$ \vvhb\ $\leq$ 2.0 mag and $\sigma$(\cnjwl) $\leq$ 0.005 mag. The blue dots denote the \cnw\ population and the red dots the \cns\ population based on the \cnjwlcor\ distribution in Figure~\ref{fig:m5}.
The \pcnjwl\ and the \pdca\ CMDs show discrete double RGB sequences, suggesting that \pdca\ can be a good surrogate \pcnjwl\ index in separating MPs.
On the other hand, the \phkctio\ and the \phkpjwl\ CMDs show broad RGB sequences, where the confusion in separating populations are severe and the transition of one population to the other is continuous.
}\label{fig:rec}
\end{figure}

Based on our populational classification from the \pcnjwl, we show various CMDs for the proper motion membership RGB stars in M5 with \vvhbmag\ in Figure~\ref{fig:cmdidx}.
In the figure, we used stars with $\sigma$(\cnjwl) $\leq$ 0.005 mag, in order to avoid confusion raised from photometric measurement uncertainties.
Note that our selection process for stars with accurate measurements does not affect the boundary between the two populations and the populational number ratio remains the same, \nrgb\ = 30:70 ($\pm$3).
It can be clearly seen that both the \cnjwl\ and the \cnpjwl\ show discrete double RGB sequences.
In these two CMDs, by definition, the \cnw\ RGB population occupies the blue sequence and the \cns\ population occupies the red sequence with well separated gaps between the two populations.
On the other hand, in the \chjwl\ CMD, the \cnw\ population occupies the red part of the sequence, meaning that the \cnw\ population has more enhanced CH abundance (i.e., carbon abundance) than the \cns\ population does.
Therefore, our results shown in Figure~\ref{fig:cmdidx} is a photometric analogue of the CN--CH anticorrelation seen in M5 RGB stars \citep[see also,][]{cohen02}.

The absence of the discreteness and the continuous transition between MPs in the \hkctio\ and the \hkpjwl\ indices deserve some attention.
As already shown in Figure~\ref{fig:filter}, the passbands of the \cactio\ and $JWL39$ filters contain the CN band at \cnwave\ and the \caii\ H and K lines, which are known to be the most dominant spectral features that affect the passbands from 3800 \AA\ to 4000 \AA\ \citep[also see Figure~1 of][]{lee15}.
If there exists no variation in the calcium abundance between MPs, which appears to be true for M5, the discrete bimodal distributions should persist both in the \hkctio\ and \hkpjwl\ indices, similar to the discrete bimodal distributions  seen in the \cnjwl\ or in the \cnpjwl.
For example, the calcium abundance of thirteen M5 RGB stars from high-resolution spectroscopy by \citet{ivans01} is [Ca/Fe] = 0.31 $\pm$ 0.04 and their results do not show any sign of a significant variation in the calcium abundance between the \cnw\ and the \cns\ RGB stars.
Therefore, contributions from elements other than calcium within the passband of the \cajwl\ filter must have erased the discreteness of the bimodal distributions in the \hkctio\ and the \hkpjwl\ CMDs.

In the \hkjwl\ CMD, it is evident that the distributions of MPs are skewed, in the sense that the \cnw\ RGB stars are preferentially located on the red part of the \hkjwl\ (i.e.\ larger values of the \hkjwl\ index at a given $V$ magnitude), and the general trend of the \hkjwl\ distribution is very similar to that of the \chjwl\ distribution, which led us to suspect that the \cajwl\ or the \caat\ filters may be contaminated by the CH molecular band, making continuous populational transitions in the \hkctio\ and in the \hkpjwl\ CMDs.

In Figure~\ref{fig:rec}, we show parallelized CMDs and histograms of RGB stars for the individual color indices.
Note that we restricted the visual magnitude range of $-$0.5 mag $\leq$ \vvhb\ $\leq$ 2.0 mag, due to the lack of the bright \cnw\ proper motion membership RGB stars with \vvhb\ $\lesssim$ $-$0.5 mag when we restricted sample RGB stars with $\sigma$(\cnjwl) $\leq$ 0.005 mag, which hinders us to achieve accurate parallelization of the two populations.
Similar to Figure~\ref{fig:cmdidx}, only the \pcnjwl\ and the \pcnpjwl\ CMDs exhibit discrete bimodal distributions with clear populational separations, while other color indices show broad spread in their colors.
In particular, the confusion in the populational tagging for individual stars is severe in the \phkctio\ and the \phkpjwl\ indices.

\begin{figure}
\epsscale{1.2}
\figurenum{7}
\plotone{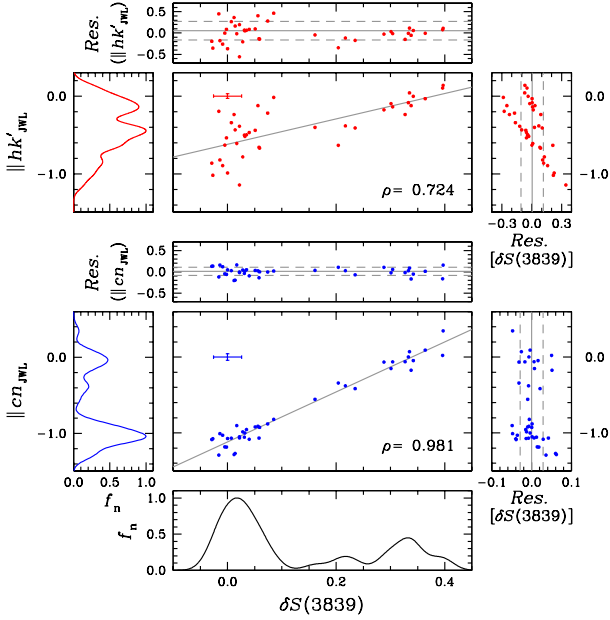}
\caption{
Plots of \ds\ versus parallelized \cnjwl\ and \hkpjwl\ for M3 RGB stars along with least-square fits.
The mean residuals ($\pm 1\sigma$) around the fitted lines are also shown with long-dashed lines.
The \phkpjwl\ index shows a weak correlation, with the correlation coefficient of $\rho$ = 0.724, against the spectroscopic \ds\ index by \citet{smolinski11}.
In sharp contrast, the plot shows that our \pcnjwl\ index is nicely correlated with the \ds, with the correlation coefficient of $\rho$ = 0.981.
We emphasize that the shape of the histogram for \pcnjwl\ is very similar as that for the \ds.
Also note that the scatter around the fitted lines are $\sigma$ = 0.217 and 0.094 for the \phkpjwl\ and the \pcnjwl, respectively, and they are at the levels of 7.2$\times\sigma$(\phkpjwl) and 2.2$\times\sigma$(\pcnjwl), where $\sigma$(\phkpjwl) and $\sigma$(\pcnjwl) are photometric measurement uncertainties.
In the right panels, we show the residuals in \ds\ around the fitted lines, finding $\sigma$ = 0.136 and 0.029 for the \phkpjwl\ and the \pcnjwl, respectively, and they are at the levels of 4.3$\times\sigma$[\ds] for \phkpjwl\ and 1.1$\times\sigma$[\ds] for \pcnjwl, where $\sigma$[\ds] is the spectroscopic measurement uncertainty by \citet{smolinski11}.
The large scatter in \phkpjwl\ is not due to the photometric or the spectroscopic measurement errors but due to its intrinsic nature as a poor CN-tracer.
}\label{fig:compcn}
\end{figure}

A more vivid example can be found in Figure~\ref{fig:compcn}, where we show plots of \ds\ versus \pcnjwl\ and \phkpjwl\ for M3 RGB stars.
Note that a plot of \ds\ versus \pcnjwl\ is already shown in our previous work \citep{lee19a}, and we show it again as a reference.
The figure shows a weak correlation between the \phkpjwl\ and \ds, with a correlation coefficient of $\rho$ = 0.724 and very large scatters around the fitted line.
We obtained the residual in \phkpjwl\ of $\sigma$ = 0.217 and it is at the level of 7.2$\times\sigma$(\phkpjwl), where $\sigma$(\phkpjwl) is the photometric measurement uncertainty.
We also calculated the residual in \ds\ around the fitted line, finding $\sigma$ = 0.136  and it is at the levels of 4.3$\times\sigma$[\ds], where $\sigma$[\ds] is the spectroscopic measurement uncertainty by \citet{smolinski11}.
The scatters of \phkpjwl\ in both axes are much larger than those of \pcnjwl,
0.094 and 0.029 for residuals in \pcnjwl\ and \ds, respectively.
The large scatter in \phkpjwl\ is not due to the photometric or the spectroscopic measurement errors but due to its intrinsic nature as a poor CN-tracer, and the same is true for the \phkctio\ index.
In particular, very large scatters around the fitted line for the \phkpjwl\ versus \ds\ can be seen in the \cnw\ populations, which is due to high carbon abundance in the \cnw\ population and, therefore, an increasing degree of the CH contamination in the \phkpjwl\ index of the \cnw\ population, as we will show later.

Finally, it should be pointed out that Figures~\ref{fig:cmdidx} and \ref{fig:rec} strongly suggest that  the skewed distributions in our \hkjwl\ and \hkpjwl (i.e., our \cajwl) are not due to the CN contamination.
If our \cajwl\ filter is contaminated by the CN molecular band, then the \cns\ population should occupy the red part of the \hkjwl\ or the \phkjwl, which is the opposite case to our results.

\begin{figure}
\epsscale{1.2}
\figurenum{8}
\plotone{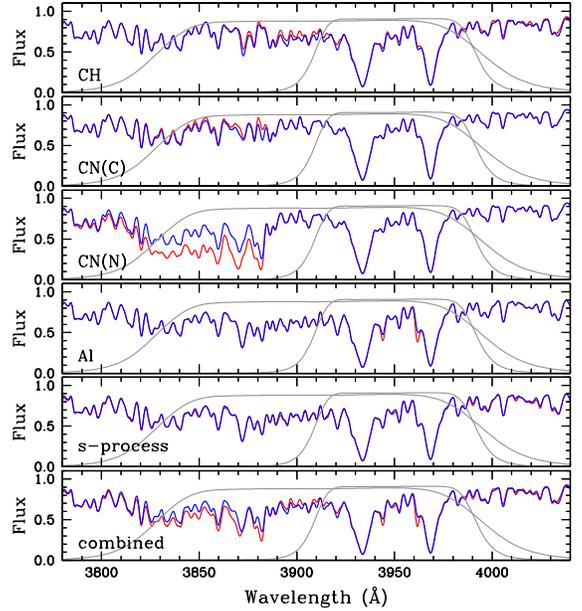}
\caption{
Comparisons of synthetic spectra of the \cnw\ (blue solid lines) and the \cns\ (red solid lines) to illustrate the contributions of individual elements.
We adopt $\Delta$[C, N, Al, $s$-process/Fe] = $-$0.8, +1.0, +1.2 and +0.5 dex. CN(C) and CN(N) denote the variations of the CN band strengths due to the variations of the carbon and the nitrogen abundances only.
The filter transmission curves for the $JWL39$ and \cajwl\ are shown with solid gray lines.
}\label{fig:synspec}
\end{figure}

\section{CH Contamination In The \caii\ $H$ and $K$ Passband}\label{s:5}
In order to understand what elements may affect the wavelength region from $\lambda$3800 to $\lambda$4000, we construct synthetic spectra of the \cnw\ and the \cns\ with varying elemental abundances using MOOG \citep{sneden73} and we show our results in Figure~\ref{fig:synspec}.
During our calculations, we assumed that the \cns\ population is depleted in the carbon and oxygen abundances by $-$0.5 and $-$0.4 dex and is enhanced in nitrogen, aluminum and $s$-process elemental abundances by 0.8, 1.0 and 0.5 dex, respectively, with respect to the \cnw\ population.
Although it does not seem to be realistic, we excluded the CN molecular lines when we construct synthetic spectra to examine the contribution from the CH molecular lines, and vice versa, for heuristic purpose.
It is not a quantitatively correct approach because the CH formation rate is also affected by the CN formation and vice versa, but our approach can render some qualitative analysis.

The top panel of Figure~\ref{fig:synspec} shows the contribution of the CH molecular lines, where the FG has stronger absorption features due to its enhanced carbon abundance and, as a result, the \hkjwl\ index value of the \cnw\ population can be slightly larger (i.e., redder) than that of the \cns\ population. 
The next two panels show the contribution of the CN molecular lines, whose formation is regulated by both the carbon and the nitrogen abundances.
The figure shows that  the CN band does not contribute in the passband of the \cajwl\ filter.
As mentioned earlier, the location of the \cnw\ population occupies the red part of the \hkjwl\ index and, therefore, it could be  an observational line of evidence that our \cajwl\ filter is free from CN contamination.
In other words, if our \cajwl\ filter is contaminated by the CN band, then, the \cns\ RGB stars would occupy the red part of the \hkjwl\ index.
The different carbon isotope ratios can also affect the CN band strengths \citep[e.g., see][]{briley89}, but they do not affect our \cajwl\ passband.

It is a well-known fact that many GCs show large star-to-star aluminum abundance variations by more than $\Delta$[Al/Fe] $\approx$ 1.0 dex, presumably resulted from the proton-capture process at high temperature or the primordial pollution by intermediate-mass AGB stars to the later generation of the stars.
The resonance lines of \ion{Al}{1} at $\lambda$3944.01 and 3961.52 are often very strong and they can affect the \hkjwl\ or the \phkjwl\ at the level of $\sigma(hk)$ $\lesssim$ 0.01 mag \citep[see also,][]{norris81,jwlnat}.
However, the influence of the aluminum lines works in opposite sense to our observations: the aluminum enhancement of the \cns\ population makes the \cns\ population redder than the \cnw\ population and, therefore, the aluminum enhancement cannot place the \cnw\ population the redder part of the \hkjwl\ sequence.

Finally, the \cns\ population is expected to have enhanced $s$-process elements, which lead the \cns\ population redder than the \cnw\ population.
Again, the variation of the $s$-process elements cannot explain the skewed \hkjwl\ or \phkjwl\ distributions.

In the synthetic spectra including all contributions from individual elements, it is evident that the \cajwl\ or the \caat\ filters, and consequently the $hk$ and \hkjwl\ indices,  contain the CH molecule lines and, as a consequence, the $hk$, \hkctio, \hkjwl, and \hkpjwl\ suffer from weak CH contamination.
However, it should be emphasized that this CH contamination is not as severe as to produce discrete bimodal RGB distributions in the $hk$ and \hkjwl\ as can be seen in M22 \citep{lee15}. 
For M5, the CH contamination works in a continuous way and its influence on the \hkjwl\ index is not significant, $\sigma$(\hkjwl) $\leq$ 0.04 mag.
On the other hand, the $hk$ split at the level of \vhb\ in M22 is about $\Delta hk$ $\geq$ 0.1 mag, which is due to the difference in the calcium abundance between the two populations with heterogeneous metallicities and the level of the populational separation in M22 is much larger than that can be seen in M5.

\begin{figure}
\epsscale{1.2}
\figurenum{9}
\plotone{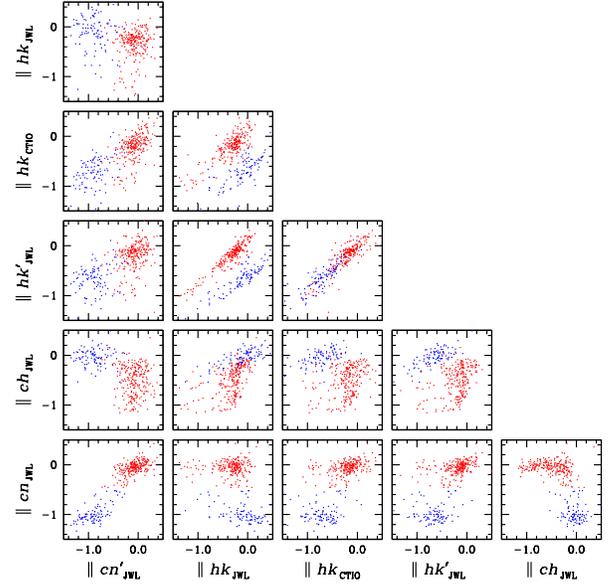}
\caption{
Comparisons of color indices of the proper motion membership M5 RGB stars with $-$0.5 $\leq$ \vvhb\ $\leq$ 2.0 mag  and $\sigma$(\cnjwl) $\leq$ 0.005 mag.
Our \pcnjwl\ is nicely correlated with the \pcnpjwl\ and is inversely correlated with our \pchjwl, a photometric analogue of the CN--CH anticorrelation of M5 RGB stars.
Note the significant spread of the $\parallel$\chjwl\ for the \cns\ population at a fixed $\parallel$\hkjwl.
}\label{fig:comp}
\end{figure}

\begin{deluxetable}{lrrrrr}
\tablenum{1}
\tablecaption{The correlation coefficients between color indices\label{tab:cor}}
\tablewidth{0pc}
\tablehead{
\multicolumn{1}{c}{Name} &
\multicolumn{1}{c}{\pcnjwl} &
\multicolumn{1}{c}{\pcnpjwl} &
\multicolumn{1}{c}{\phkjwl} &
\multicolumn{1}{c}{\phkctio} &
\multicolumn{1}{c}{\phkpjwl}
}
\startdata
\pcnpjwl &    0.918 &          &        & & \\
\phkjwl  & $-$0.319 & $-$0.247 &        & & \\
\phkctio &    0.676 &    0.741 &  0.384 & &   \\
\phkpjwl &    0.651 &    0.633 &  0.507 &    0.917 & \\
\pchjwl  & $-$0.586 & $-$0.502 &  0.575 & $-$0.135 & $-$0.072 \\
\enddata 
\end{deluxetable}

In Figure~\ref{fig:comp} and Table~\ref{tab:cor}, we show comparisons of color indices and correlation coefficients between various color indices. Our \pcnjwl\ is nicely correlated with the \pcnpjwl. Therefore, the \cnpjwl\ index makes a good surrogate \cnjwl\ index.
It is interesting to note that \pcnjwl\ is inversely correlated with our \pchjwl, but with a step-function wise, a photometric analogue of the CN--CH anticorrelation of M5 RGB stars, as we already mentioned.
Our plot of the \pcnjwl\ versus \pchjwl\ shows that the evolutions of the CN and CH between the \cnw\ and \cns\ populations are not continuous. Apparently, our photometric CH abundance of the \cnw\ population does not vary, while that of the \cns\ population shows a significant variation.

\bigskip
\emph{The lesson learned from our exercise is that neither the \hkctio\ nor the \hkpjwl, subsequently neither the \phkctio\ nor the \phkpjwl, can serve as an accurate population tracers in normal GCs without any metallicity spread.
Only our \cnjwl\ and, perhaps, the \cnpjwl, subsequently the \pcnjwl\ and the \pcnpjwl, are the most reliable population tracers.}

\begin{deluxetable*}{ccccccccccc}
\tablenum{2}
\tablecaption{Integration times (s) for NGC\,6723\label{tab:obs}}
\tablewidth{0pc}
\tablehead{
\multicolumn{1}{c}{} &
\multicolumn{5}{c}{CTIO Filters} &
\multicolumn{1}{c}{} & \multicolumn{4}{c}{New Filters} \\
\cline{2-6}\cline{8-11}
\colhead{} &
\colhead{$y$} & \colhead{$b$} & \colhead{$v$} & \colhead{$u$} & 
\colhead{$Ca_{\rm CTIO}$} & \colhead{} & 
\colhead{$y$} &
\colhead{$b$} & 
\colhead{$Ca_{\rm JWL}$} &
\colhead{$JWL39$}
}
\startdata
NGC\,6723 & 1910 & 3340 & 2700 & 1800 & 11500 & & 4450 & 10680 & 44420 & 10200  \\ 
\enddata 
\end{deluxetable*}

\section{Multiple Stellar Populations in NGC\,6723}\label{s:6}
Here, we present the multiple stellar population study of NGC\,6723 based on our \cnjwl\ and \cnpjwl\ indices.

\subsection{Observations}
The observations for NGC\,6723 were performed over 15 nights in seven runs from 2007 September to 2014 May using the CTIO 1.0 m telescope, which was equipped with a STA 4k $\times$ 4k CCD camera, providing a plate scale of 0\farcs289 pixel$^{-1}$ and  a field of view (FOV) of about 20\arcmin\ $\times$ 20\arcmin.
The effective FOV of the final combined image of NGC\,6723 is about 35\arcmin\ $\times$ \ 27\arcmin, which is at least 8 times larger than that of the half-light radius of the cluster, 1\farcm53 \citep[][updated as in 2010]{harris96}.
We show the total integration times for the cluster in Table~\ref{tab:obs}.

The raw data handling were described in detail in our previous works \citep{lee14,lee15,lp16}. The photometry of NGC\,6723 and standard stars were analyzed using DAOPHOTII, DAOGROW, ALLSTAR and ALLFRAME, and  COLLECT-CCDAVE-NEWTRIAL packages \citep{pbs87,pbs94,lc99}.
We performed artificial star experiments \citep[e.g., see][]{sh88,lee15} and obtained the completeness fractions, $f_{\rm complete}$, as a function of $V$ magnitude, whose utility will be discussed more fully later.

We calculated the astrometric solutions for individual stars using the data extracted from the Naval  Observatory Merged Astrometric Dataset \citep[NOMAD,][]{nomad} and the IRAF IMCOORS package.

\begin{figure}
\epsscale{1.1}
\figurenum{10}
\plotone{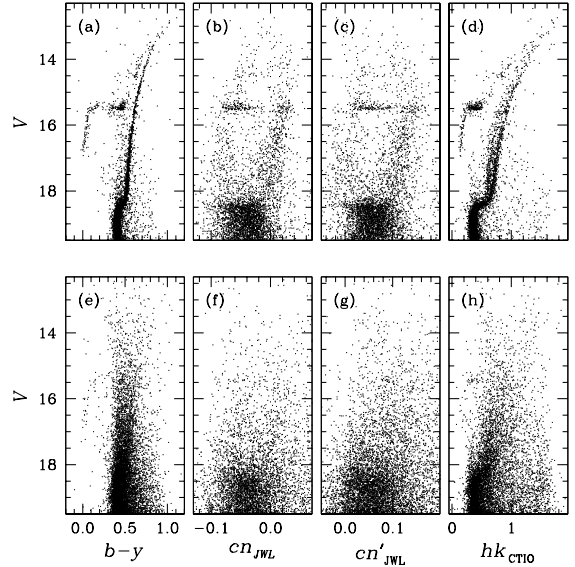}
\caption{
(a)--(d) CMDs of the NGC\,6723 membership stars based on the proper motion study of the second \gaia\ data release.
(e)--(h) CMDs of the off-cluster stars toward NGC\,6723 field.
}\label{fig:n6723cmd}
\end{figure}

\subsection{Color--Magnitude Diagrams}
As we showed in our previous work \citep{lee14}, NGC\,6723 is an intermediate metallicity GC located toward the Galactic bulge ($l$ = 0\fdg07, $b$ = $-$17\fdg30) and is heavily contaminated by off-cluster field stars.
The recent study of the space orbit of NGC\,6723 revealed that its apogalactic distance is less than 3 kpc \citep{baumgardt19}, making it a true inner Galactic GC.
Using the second \gaia\ date release \citep{gaiadr2} and our multi-color photometry \citep[see, e.g.,][for the versatility of multi-color photometry]{lee15}, we removed the off-cluster field stars \citep[e.g., see also][]{bastian18,marino18,milone18}.  We show selected CMDs for NGC\,6723 in Figure~\ref{fig:n6723cmd}, where discrete double RGB sequences can clearly be seen in the \cnjwl\ and \cnpjwl\ CMDs, while a hint for the double RGB populations can be noticed in the \hkctio\ CMD.

\begin{figure}
\epsscale{1.1}
\figurenum{11}
\plotone{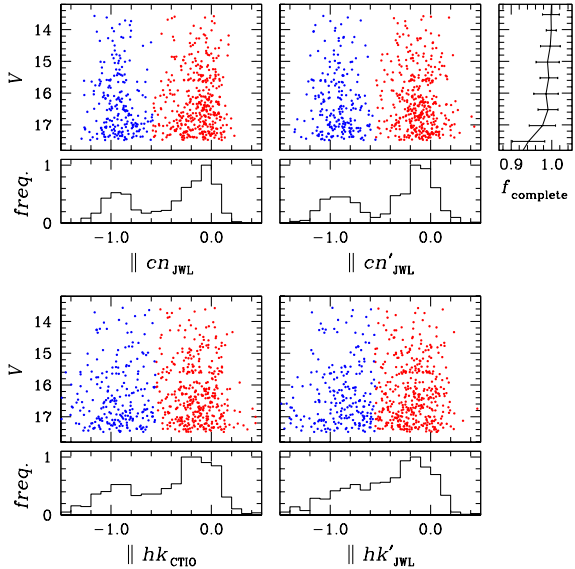}
\caption{
Parallelized CMDs and distributions for NGC\,6723 RGB stars. The discrete double RGB sequences can be clearly seen in the \pcnjwl\ and the \pcnpjwl\ indices, while rather continuous transitions from one population to the other in the \phkctio\ and the \phkpjwl\ indices.
In each panel, the blue and the red colors indicate individual populations based on each color index as presented in Table~\ref{tab:pop}.
We also show completeness fractions, $f_{\rm complete}$, at each magnitude bin estimated from the artificial star experiments.
}\label{fig:n6723rec}
\end{figure}

\begin{deluxetable}{lcc}
\tablenum{3}
\tablecaption{NGC\,6723 RGB populational number ratios from various color indices\label{tab:pop}}
\tablewidth{0pc}
\tablehead{
\multicolumn{1}{c}{Color Index} &
\multicolumn{1}{c}{\nrgb} &
\multicolumn{1}{c}{$\sigma$}
}
\startdata
\pcnjwl & 35.5:64.5  & 2.8 \\
\pcnpjwl & 35.5:64.5  & 2.8 \\
\phkctio & 37.4:62.6  & 2.9 \\
\phkpjwl & 37.5:62.5  & 2.9 \\
\enddata 
\end{deluxetable}

\subsection{Populational Ratios of Red-Giant Branch Stars}
We parallelized the selected color indices using Eq. (\ref{eq:pl}) and we show the parallelized CMDs and histograms in Figure~\ref{fig:n6723rec}, where we show the \cnw\ and \cns\ RGB stars with \vvhbmag\ classified from individual indices. Same as Figure~\ref{fig:rec}, the \pcnjwl\ and \pcnpjwl\ CMDs show discrete RGB populations, while the \phkctio\ and \phkpjwl\ CMDs show rather continuous transitions from one population to the other and severe confusion in the populational tagging for individual stars.

In Table~\ref{tab:pop}, we show the RGB populational ratios from four color indices using the EM estimator that we already explained in \S\ref{s:4}. Our results show that the overall populational number ratios from individual color indices are in good agreement with each other, $\approx$ 36:64 ($\pm$3). At the same time, our RGB populational number ratios are in excellent agreement with that of \citet{milone17}, who obtained the FG fraction of 0.363 $\pm$ 0.017 in the central part of the cluster, which is not a surprise because the RGB populational number ratios do not show any radial gradient in NGC\,6723 as we will show later.
On the other hand, our RGB number ratios from the \pcnjwl\ or \pcnpjwl\ indices are slightly different from that by \citet{lim16}, who obtained \nrgb\ = 38:62, but both results are in agreement within statistical errors.
In fact, the result by \citet{lim16} is more close to those of our results from the \phkctio\ or \phkpjwl.

The populational tagging for individual stars, however, can be somewhat different. In the \pcnpjwl\ CMD, a weak confusion in the lower RGB stars can be seen, in particular a few number of the \cnw\ stars identified from the \pcnjwl\ index occupy the \cns\ regime in the \pcnpjwl. The situations are even worse in the \phkctio\ and \phkpjwl\ CMDs, where considerable numbers of stars occupy the counterpart regime, making the populational tagging for the individual stars can be unreliable for these two color indices.

\begin{figure}
\epsscale{1.1}
\figurenum{12}
\plotone{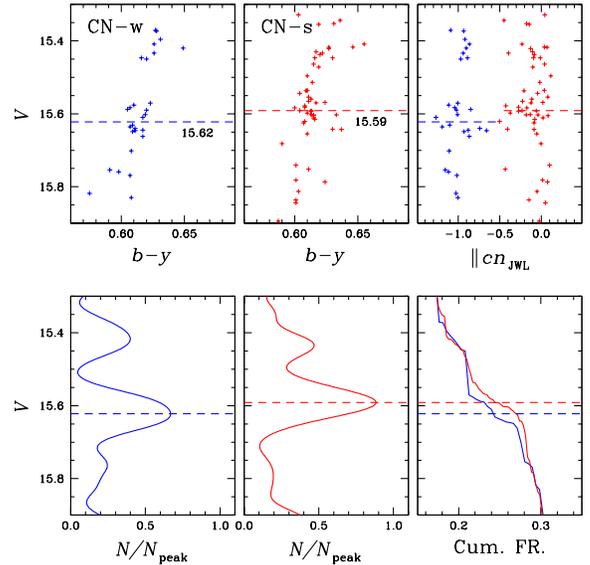}
\caption{
(Top panels) CMDs of the \cnw\ and \cns\ RGB stars in NGC\,6723. The dashed lines denote the RGB bump $V$ magnitude, \vbump.
(Bottom panels) Generalized differential and cumulative LFs.
The mean RGB bump of the \cns\ population is 0.031 $\pm$ 0.030 mag brighter than that of the \cnw\ population.
}\label{fig:n6723bump}
\end{figure}

\subsubsection{Red-Giant Branch Bumps}
In GC RGB stars, the helium abundance is very difficult to measure due to the lack of the absorption lines in the visual passband. Direct spectroscopic measurements can be possible using the chromospheric \ion{He}{1} absorption line at $\lambda$10830\AA, but the result from this line is sensitively dependent on the structure and dynamics of the stellar atmosphere \citep[e.g., see][]{dupree13}.
For blue HB (BHB) stars cooler than the Grundahl jump \citep{grundahl99}, the helium abundance can be measured from the photospheric \ion{He}{1} lines at $\lambda$5876\AA. However, due to difficulties involved in the absorption strength measurements in fainter GC BHB stars and in the model atmosphere constructions, the inferred helium abundance from this absorption line can be incorrect. For example, \citet{gratton15} obtained very high mean helium abundance ($Y$ $\approx$ 0.37) from the two BHB stars in NGC\,6723, and they cautioned about using their helium measurement.

Alternatively, one can rely on the indirect photometric methods using the helium sensitive characteristics during the low-mass stellar evolutions, such as the RGB bump (RGBB) luminosity \citep[e.g., see][and references therein]{cassisi13,lee15,lee17,lee18,milone15,largioia18}.
When the very thin H-burning shell crosses the discontinuity in the chemical composition and lowered mean molecular weight left by the deepest penetration of the convective envelope during the ascent of the RGB, the RGB stars experience slower evolution and temporary drop in luminosity, the so-called RGBB. The luminosity of the RGBB depends on metallicity, stellar mass (hence age), and helium abundance. At a given age and metallicity, the RGBB luminosity increases with helium abundance.

We compared the magnitude difference of the RGBB between the two populations and we show our results in Figure~\ref{fig:n6723bump}. We found that the mean RGBB magnitude of the \cns\ population is 0.031 $\pm$ 0.030 mag brighter than that of the \cnw\ population. If there exist no differences in metallicity and age between the two RGB populations, the difference in the RGBB magnitude can translate into the difference in the mean helium abundance of $\Delta Y$ = 0.012 $\pm$ 0.012, in the sense that the \cns\ population is slightly more helium enhanced than the \cnw\ population is. Our result is in good agreement with that of \citet{milone18}, who obtained the average helium difference between two generations of the stars in NGC\,6723 of 0.005 $\pm$ 0.006 from their \hst\ photometry. Our finding is also consistent with the widely accepted idea that the \cns\ population in GCs are helium enhanced population \citep[e.g., see][]{lee17,lee18,largioia18,milone18}.

\begin{figure}
\epsscale{1.1}
\figurenum{13}
\plotone{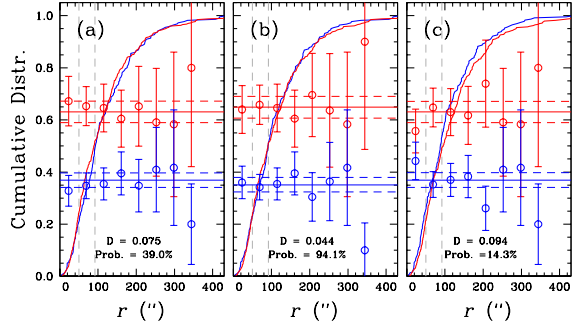}
\caption{
(a) A comparison of the cumulative radial distributions of the \cnw\ (blue) and the \cns\ (red) RGB stars classified from the \pcnjwl\ index in NGC\,6723. The blue and the red horizontal lines denote the mean fractions for each population with $\pm1\sigma$.
(b) Same as (a), but from \pcnpjwl.
(c) Same as (a), but from \phkctio.
}\label{fig:n6723rad}
\end{figure}

\begin{figure}
\epsscale{1.1}
\figurenum{14}
\plotone{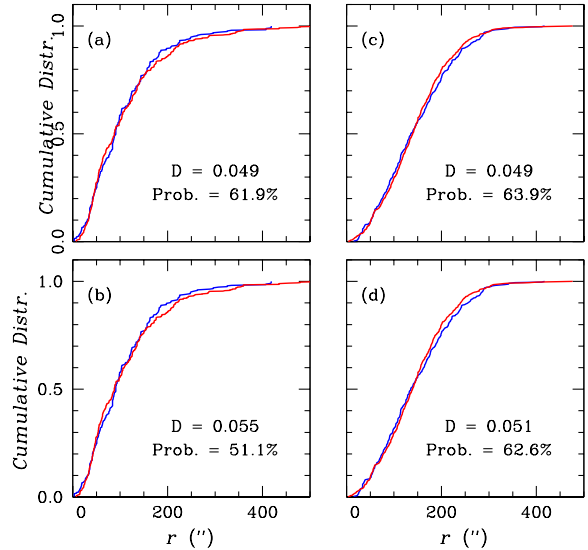}
\caption{
Mote Carlo simulations of the artificial radial distributions of the two populations in NGC\,6723.
(a) Using empirical radial distributions as shown in Figure~\ref{fig:n6723rad} with the populational number ratio of \nrgb\ = 35.5:64.5.
(b) Same as (a) but applying completeness fractions as shown in Figure~\ref{fig:n6723rec}. 
(c) Same as (a), but using Gaussian radial distributions.
(c) Same as (b), but using Gaussian radial distributions.
Our simulations show that a slightly incomplete detection of stars does not affect the radial distributions.
}\label{fig:radMC}
\end{figure}

\begin{deluxetable}{cc}
\tablenum{4}
\tablecaption{Probability for being $p$-value of less than 0.002 (i.e., different radial distributions for the two populations)\label{tab:random}}
\tablewidth{0pc}
\tablehead{
\multicolumn{1}{c}{Sampling domains} &
\multicolumn{1}{c}{$P$($p$-value $\leq$  0.002)} 
}
\startdata
Empirical & 4.46\% \\
Empirical\tablenotemark{1} & 0.18 \% \\
Gaussian & 0.20 \% \\
\enddata 
\tablenotetext{1}{Sampling from the whole RGB stars.}
\end{deluxetable}

\subsubsection{Radial Distributions}
As we mentioned earlier, \citet{lim16} reported that the radial distribution of the \cnw\ RGB stars is more centrally concentrated than that of the \cns\ with a $p$-value of 0.002 for being drawn from same parent distributions. Then, they claimed that the mass segregation is responsible for the difference in the radial distributions.

In Figure~\ref{fig:n6723rad}, we show cumulative radial distributions from the \pcnjwl, \pcnpjwl, and \phkctio. Note that our \hkctio\ is a exactly same index as $hk_{Ca+CN}$ by \citet{lim16}. As shown in the figure, the radial distributions for the \cnw\ and \cns\ populations from all three color indices are almost identical. Our Kolmogorov--Smirnov (K--S) tests show that the \cnw\ and \cns\ populations are most likely drawn from the same parent distributions regardless of the color indices. 
At the same time, the populational ratios from the \pcnjwl\ and \pcnpjwl\ indices remain flat against the radial distance from the center.

It should be emphasized that, despite the severe confusion in populational tagging in the \phkctio\ index as shown in Figures~\ref{fig:rec} and \ref{fig:comp}, the apparently almost identical radial distributions between the two populations in the \phkctio\ index does not imply that the \phkctio\ index performs populational tagging in a correct way, but it is due to the natural consequence of the absence of the radial sensitivity in the distribution of the two populations.
We constructed both the fully empirical and the analytical models to perform Monte Carlo simulations to test this idea \citep[see][]{lee15,lee19b}. 
Note that our empirical models use our observed \cnw\ and \cns\ distributions, the populational number ratio and luminosity functions.
In Table~\ref{tab:random}, we show the results from our randomization tests for probability of having the $p$-value returned from our K--S tests less than 0.002, which is the one that \citet{lim16} claimed.
Our tests strongly suggest that there exists no difference in radial distributions between the two RGB populations, which is consistent with that of \citet{gratton15}, who compared the cumulative radial distribution of extreme-BHB stars (i.e., the \cns\ population) and other stars on the HB (i.e., the \cnw\ population) in NGC\,6723 using our previous $BV$ photometry \citep{lee14}, finding no evidence of different radial distributions between the two HB populations.

An incomplete detection of GC RGB stars does not affect our results for the radial distributions and populational number ratios.
Using models constructed above, we performed Monte Carlo simulations and we show our results in Figure~\ref{fig:radMC}.
The results from our fully empirical Monte Carlo simulations presented in Figure~\ref{fig:radMC}(a) show almost identical radial distributions for both populations similar to that already presented in Figure~\ref{fig:n6723rad}(a).
In order to simulate our potential incomplete detection of NGC\,6723 RGB stars, by applying the completeness fractions, $f_{\rm complete}$, we calculated probability of being detected in our Monte Carlo simulations. 
We show our results in Figure~\ref{fig:radMC}(b), where the similar radial distributions for both populations persist.
Note that our approach applying the completeness fractions to the observed radial distributions is not an exactly correct way to realize the effect of incomplete detection on the radial distributions. 
In fact, the inverse process of our approach should be sought, but it turned out to be very delicate to construct such models.
Although it is not exactly correct, very similar results returned from our simulations strongly suggest that any effect arisen from the incomplete detection of stars on the radial distribution, and furthermore on the populational number ratios, can be negligible.

It is very clear that our results are significantly different from those of \citet{lim16} and we strongly believe that the results presented by \citet{lim16} are in error, due to either their incorrect populational assignment of the individual stars based on the \hkctio\ index, a rather poor CN tracer or their incorrect photometric measurements.

\begin{figure}
\epsscale{1.1}
\figurenum{15}
\plotone{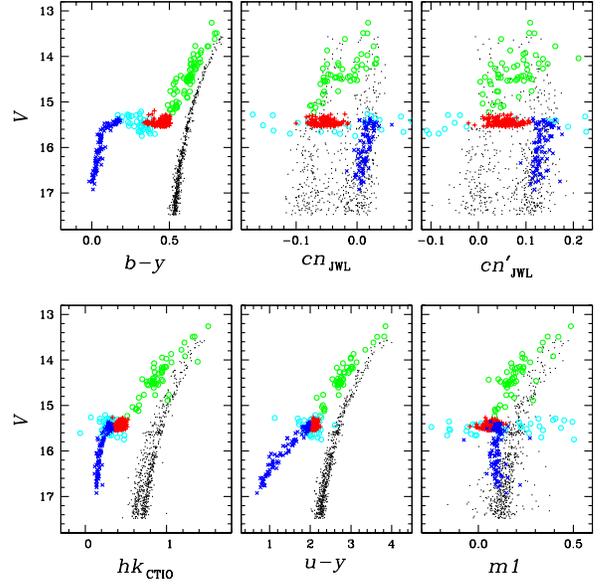}
\caption{
CMDs of bright stars in NGC\,6723. Black dots, green circles, red plus signs, blue crosses, and cyan circles denote the RGB, AGB, red HB (RHB), BHB, and RRL, respectively, where the photometry for the RRL variables is not phase-averaged.
The \hkctio\ CMD shows that the \hkctio\ index is not suitable to study the MPs of the HB and AGB stars. 
Note that the BHB should not be considered as the \cns\ population based on their \cnjwl\ and \cnpjwl\ indices. The surface temperature of the BHB stars are hot enough to suppress the CN band formation in their atmospheres.
}\label{fig:n6723hbagb}
\end{figure}

\begin{figure}
\epsscale{1.2}
\figurenum{16}
\plotone{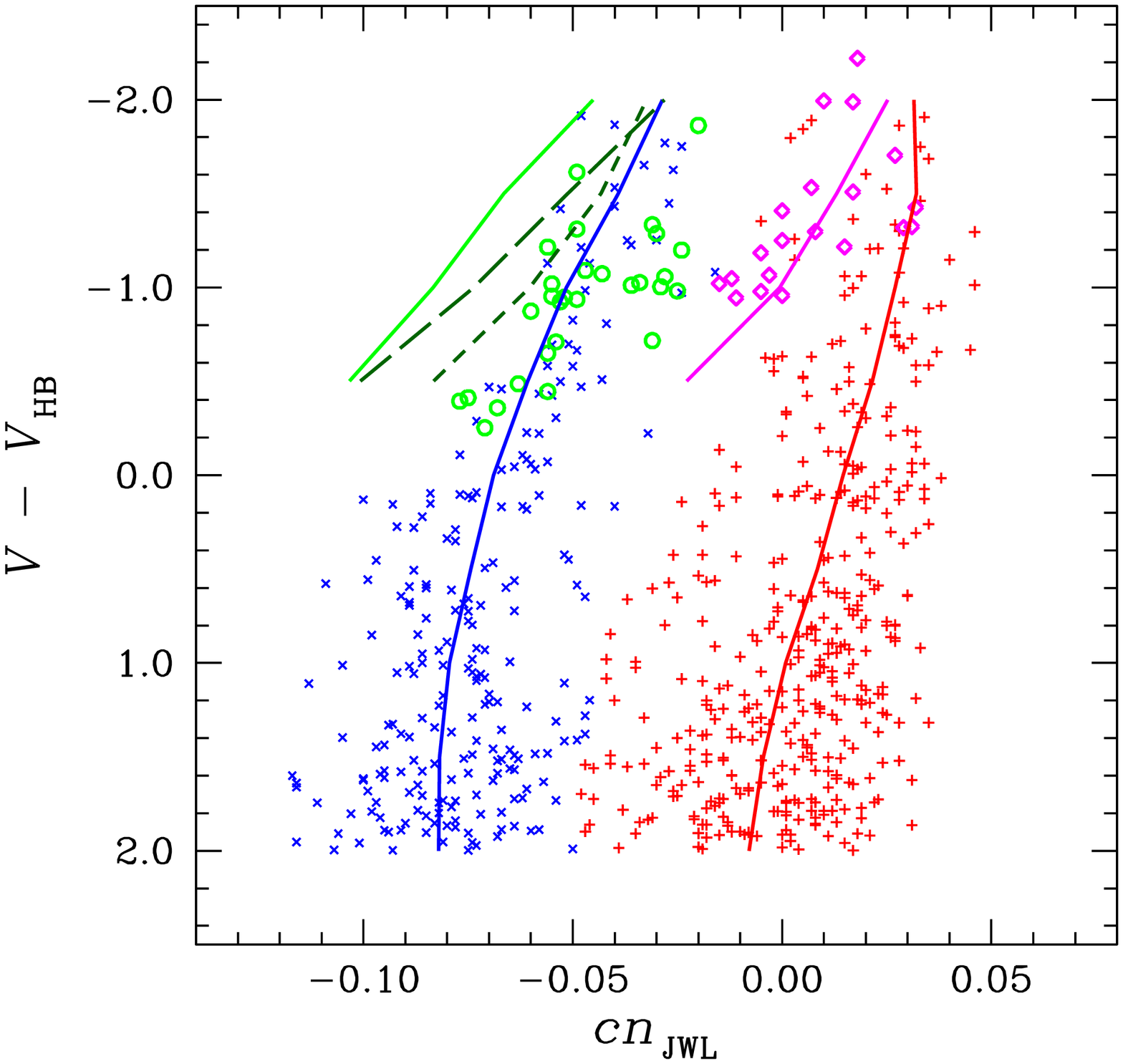}
\caption{
The CMD of RGB stars with \vvhbmag, and AGB stars in NGC~6723. The \cnw\ and the \cns\ RGB stars are denoted with blue crosses and red plus signs, while the \cnw\ and the \cns\ AGB stars with green circles and magenta diamonds.
We also show synthetic \cnjwl. The blue and the red solid lines show the synthetic \cnjwl\ for the \cnw\ and \cns\ RGB, while the green and the magenta solid lines show the synthetic \cnjwl\ for the \cnw\ and \cns\ AGB.
Note that the synthetic \cnjwl\ index for the \cnw\ AGB stars does not match with the observations.
The dark-green long-dashed line denotes the synthetic \cnjwl\ AGB sequence for \ciso\ $\approx$ 2.3 (assuming initial \ciso\ $\approx$ 20) and the dark-green dashed line shows that for the enhanced nitrogen abundance by 0.1 dex.
}\label{fig:n6723syncmd}
\end{figure}

\begin{figure}
\epsscale{1.2}
\figurenum{17}
\plotone{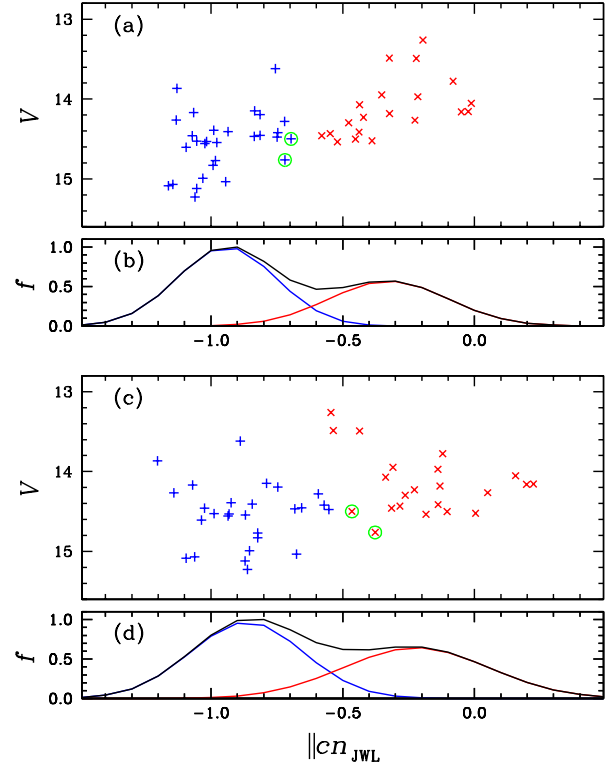}
\caption{
(a) The \pcnjwl\ CMD of AGB stars in NGC\,6723, using the RGB fiducials to derive the \pcnjwl. We show green circles for the two \cnw\ AGB stars, from the RGB fiducials, which turn out to be \cns\ AGB stars from the AGB fiducials in (c).
(b) The histogram of the AGB populations.
(c) Same as (a) but using the AGB fiducials to derive the \pcnjwl. Two red crosses encircled by green circles indicate the \cns\ AGB stars from the AGB fiducials but the \cnw\ AGB stars from the RGB fiducials as shown in (a).
}\label{fig:n6723agbpop}
\end{figure}

\begin{deluxetable}{ccc}
\tablenum{5}
\tablecaption{AGB populational number ratios\label{tab:agbpop}}
\tablewidth{0pc}
\tablehead{
\multicolumn{1}{c}{} &
\multicolumn{1}{c}{\nrgb} &
\multicolumn{1}{c}{$\sigma$} 
}
\startdata
RGB    & 35.5:64.5  & ~2.8 \\
Case 1 & 59.2:40.8  & 13.9 \\
Case 2 & 53.7:46.3  & 13.0 \\
Case 3 & 45.9:54.1  & 13.5 \\
Case 4 & 42.1:57.9  & 12.7 \\
\enddata 
\noindent Case 1: Using the RGB fiducials for the populational separation.\\
\noindent Case 2: Using the AGB fiducials for the populational separation.\\
\noindent Case 3: From bright AGBs with $V$ $\leq$ 14.54 mag for Case 1.\\
\noindent Case 4: From bright AGBs with $V$ $\leq$ 14.54 mag for Case 2.\\
\end{deluxetable}

\subsection{Asymptotic Giant Branch Stars}
We also explore the AGB population in NGC\,6723. As we already demonstrated for M5 and NGC\,6752 \citep{lee17,lee18}, the effective temperature of the AGB stars are not hot enough to suppress the CN formation and, therefore, our \cnjwl\ index can be a very powerful tool to study the AGB populations in GCs \citep[see also the work by][]{marino17}.

In Figure~\ref{fig:n6723hbagb}, we show the CMDs for AGB stars in NGC\,6723, along with the RGB and HB stars, where one can clearly see that our \cnjwl\ and, perhaps, \cnpjwl\ indices are well-suited for studying AGB stars, while other indices including \hkctio, $u-y$ and $m1$ suffer from confusion in the AGB populational tagging and they may not be useful.
In our \cnjwl\ and \cnpjwl\ CMDs, the positions of the BHB stars happened to be overlapped with those of the \cns\ RGB stars, although the BHB stars in NGC\,6723 are most likely belonging to the helium enhanced population of the cluster \citep{gratton15} and, therefore, the progeny of the \cns\ RGB. 
The fact is that, the effective temperature of the BHB stars are too high and, therefore, the CN molecule formation is suppressed in their atmospheres and the influence of the hydrogen lines are more important in some of the color indices of the BHB stars.

\subsubsection{Synthetic population models and the need for extra abundance variation in the \cnw\ AGB stars}
In order to understand the behavior of the individual populations on the \cnjwl\ CMD, we constructed synthetic population models.
Assuming [Fe/H] $\approx$ $-$1.0 dex for NGC\,6723\footnote{Recently, \citet{crestani19} obtained [Fe/H] = $-$0.93 $\pm$ 0.05 from 11 RGB stars.}, we adopted the stellar evolution isochrones and the HB/AGB evolution tracks from the Dartmouth Stellar Evolution Database \citep{dotter08}. Then, we used ATLAS12 and SYNTHE to calculate high resolution spectra with varying elemental abundances for both populations \citep{kurucz70,kurucz05,sbordone04,castelli05}.
The carbon and nitrogen abundances for individual populations in NGC\,6723 are not known and we simply adopted those values for M5 \citep{cohen02}:
we assumed [C/Fe] = $-$0.2, [N/Fe] = $-$0.1 and [O/Fe] = +0.3 dex for the \cnw\ population, while [C/Fe] = $-$0.8, [N/Fe] = +1.3 and [O/Fe] = $-$0.2 dex for the \cns\ populations.

In Figure~\ref{fig:n6723syncmd}, we show our results. 
Our synthetic RGB fiducial sequences trace the observed data with great satisfaction, and so does the \cns\ synthetic AGB fiducial sequence.
However, the synthetic fiducial sequence for the \cnw\ AGB population appears to be problematic. Our synthetic \cnjwl\ index is about 0.035 mag smaller than our photometry.
We calculated the synthetic \cnw\ RGB and AGB fiducial sequences with various combinations of elemental abundances, but our calculations always failed to trace both the RGB and the AGB stars simultaneously.
Assuming our adopted CNO abundance for the \cnw\ population is not in great error, there are at least two possibilities that make the synthetic \cnw\ AGB fiducial sequence redder in the \cnjwl\ index: (1) The nitrogen enhancement and (2) the decrements in the \ciso\ ratio \citep[e.g., see][]{briley89}.

Our \cnjwl\ index is an accurate measure of the CN band strengths at \cnwave, which is mostly governed by the nitrogen abundance, because nitrogen is less abundant than carbon.
Using ATLAS12 and SYNTHE, we calculate the model atmospheres and the synthetic spectra for the enhanced nitrogen abundance by  $\Delta$[N/Fe] = 0.1 dex. 
As shown in the figure, the nitrogen enhancement can contribute to our \cnjwl\ index by $\approx$ 0.02 mag.

To investigate the effect of the variations in the \ciso\ ratios onto our \cnjwl\ index,
we used MOOG to generate the synthetic spectra.
The decrement of the \ciso\ ratio from $\approx$ 20 to $\approx$ 2.5 can result in $\Delta$\cnjwl\ $\approx$ +0.003 mag for the \cnw\ AGB stars at \vvhb\ = $-$0.5 mag and $\approx$ +0.016 mag at \vvhb\ = $-$2.0 mag.

It should be emphasized that both the nitrogen enhancement and the decrement in the \ciso\ ratio will have the effect of a similar degree on the RGB sequences in both populations. Therefore, an extra variation in elemental abundances is required only for the \cnw\ AGB population to match the synthetic fiducial sequences with observations.

Finally, we also considered the effect of the variation of the helium abundance, finding that the enhanced helium abundance does not affect our \cnjwl\ index. We calculated the synthetic spectra with enhanced helium abundance by $\Delta Y$ $\approx$ 0.1 and found that the difference in the \cnjwl\ index between the normal and the enhanced helium spectra is no larger than $\Delta$\cnjwl\ = 0.001 mag, i.e., a null effect.

Recently, \citet{maas19} reported carbon isotope ratios in giants stars in the globular cluster M10, finding that the decrement in the \ciso\ ratios both in the \cnw\ and the \cns\ populations\footnote{Note that the \cnw\ and the \cns\ populations in our study correspond to the CN-Normal and the CN-Enhanced population by \citet{maas19}.} in M10.
They discuss further that to match their measurements of the surface carbon and carbon isotope ratios requires two different thermohaline mixing efficiencies.
If our results presented in this study are correct, a more complicated mixing scenario would be required to explain our results.

\subsubsection{AGB populational number ratio}
It has long been noticed that the lighter elemental abundances in AGB stars are different from those of the RGB stars in some GCs \citep[e.g., see][]{sneden00,campbell13}.
At the same time, the AGB populational number ratio can be significantly different from that of the RGB in GCs, due to the different evolutionary paths of the AGB stars with different elemental abundances, especially helium, and masses \citep[e.g., see Figure~20 of][]{lee17}.

In Figure~\ref{fig:n6723syncmd}, the lack of faint \cns\ AGB stars is evident in NGC\,6723, reminiscent of the NGC\,6752 AGB populations \citep{campbell13,lee18}, which is consistent with the idea that stellar evolutionary paths of the BHB stars, presumably the helium enhanced SG population, do not pass through the lower part of the AGB sequence.
To test this idea, we derived the AGB populational number ratios and we show our results in Table~\ref{tab:agbpop}.

First, we derived the \pcnjwl\ using the RGB fiducials and applied the EM estimator to calculate the AGB populational number ratio, obtaining \nrgb\ = 59:41 ($\pm$14), significantly different from that of the RGB, 35.5:64.5 ($\pm$2.8).
Next, we also calculated the AGB number ratio using the AGB fiducials. As we already showed in Figure~\ref{fig:n6723syncmd}, the slopes of the AGB populations appear to be slightly different from those of the RGB populations in our \cnjwl\ CMD, which will require slightly different fiducials in the parallelization processes between the RGB and the AGB stars.
Hence, we derived the AGB fiducial sequences and calculated the AGB populational number ratio, finding \nrgb\ = 54:46 ($\pm$13), which is still significantly different from that of the RGB.
As shown in Figure~\ref{fig:n6723agbpop}, only two AGB stars suffer from confusion in their populational tagging from different parallelizations.
Our results strongly suggest that the adopted AGB fiducial may not be the source of the discrepancy in the populational number ratios between the RGB and the AGB in NGC\,6723.

If we consider the bright AGB stars with $V$ $\leq$ 14.54 mag only, where both the \cnw\ and \cns\ AGB stars exist, the AGB number ratio becomes \nrgb\ = 46:54 ($\pm$14) from the RGB fiducials and 42:57 ($\pm$13) from the AGB fiducials. Both results are in agreement with that of the RGB within the statistical errors.
It is believed that a significant fraction of the BHB stars (i.e., presumably the \cns\ HB populations with enhanced helium contents) may have evolved into the \agbm\ stars, which may be responsible for the rather larger fraction of the \cnw\ AGB population in NGC\,6723.

\section{Summary}
In this work, we showed that \cnpjwl\ (= $Ca_{\rm CTIO} - Ca_{\rm JWL}$) can be an excellent surrogate \cnjwl\ index and an excellent populational tagger in the study of the MPs in GCs.

We also showed the CH contamination in the passband of \ion{Ca}{2} H \& K region and the sidebands of the CH G band, which can add a small, but non-negligible amount of uncertainties to the measurements.

We introduced a new color index, \chjwl\ [=$(JWL43 - b) - (b-y)$], which can measure accurate CH G band strengths, hence the carbon abundance. Our photometric CN--CH relation of the large sample of the M5 RGB stars showed that the evolutions of the CN and CH between the \cnw\ and \cns\ populations are not continuous. The CH abundance of the \cnw\ population does not appear to vary, while a significant variation can be seen in that of the \cns\ population.

Armed with our new color indices, we investigated the MPs of NGC\,6723. 
Using our \cnjwl\ and \cnpjwl\ indices, we obtained the RGB populational number ratio of \nrgb\ = 35.5:64.5 ($\pm$2.8), in excellent agreement with that of \citet{milone17}, who obtained the fraction of the first generation of stars of 0.363 $\pm$ 0.017.
In sharp contrast to the result by \citet{lim16}, who argued a reversed cumulative radial distribution in NGC\,6723 (i.e., the more centrally concentrated \cnw\ RGB population), all of our cumulative radial distributions from various color indices showed the similar cumulative radial distributions between the two RGB populations in NGC\,6723. Our statistical tests showed that possibilities for having the results claimed by \citet{lim16} is less than 5\%, i.e.\ highly improbable. It is strongly believed that the results by \citet{lim16} were in error due to their inaccurate populational assignment from their \hkctio\ index, a poor populational tagger.

Similar to normal GCs with distinctive MPs, the helium abundance of the \cns\ population inferred from the RGBB magnitude appears to be slightly enhanced by about $\Delta Y$ = 0.012 $\pm$ 0.012.

We constructed synthetic population models to quantitatively investigate the individual populations, finding that no models with bimodal elemental abundances can trace the whole RGB and AGB populations simultaneously. In particular, the \cnw\ AGB population turned out to be problematic. To match the \cnw\ AGB stars within the current framework may require a mild nitrogen enhancement and/or a large decrement in the \ciso\ ratio in the \cnw\ AGB stars. On the other hand, the effect of the helium enhancement is nil in our \cnjwl\ index.

Finally, our \cnjwl\ and \cnpjwl\ CMDs clearly showed the discrete double AGB populations in NGC\,6723, whose bright AGB populational number ratio is in marginally agreement with that of the RGB stars within the statistical errors.
The lack of the faint \cns\ AGB stars may indicate that a significant fraction of the \cns\ HB stars (presumably BHB stars with enhanced helium contents) must have evolved into the \agbm\ phase and failed to reach the AGB phase.

\acknowledgements
J.-W.L.\ acknowledges financial support from the Basic Science Research Program (grant no. 2016-R1A2B4014741) through the National Research Foundation of Korea (NRF) funded by the Korea government (MSIP). He also thanks Prof.\ Sneden for his kind discussion on the \ciso\ ratios and the anonymous referee for useful comments.


\begin{thebibliography}{}

\bibitem[Anthony-Twarog et al.(1991)]{att91}
Anthony-Twarog, B.\ J., Laird, J.\ N., Payne, D., \& Twarog, B.\ A.\ 1991, \aj, 101, 1902

\bibitem[Bastian et al.(2018)]{bastian18}
Bastian, N., Kamann, S., Cabrera-Ziri, I., et al.\ 2019, \mnras, 480, 3739

\bibitem[Baumgardt et al.(2019)]{baumgardt19}
Baumgardt, H., Hilker, M., Sollima, A., \& Bellini, A.\ 2019, \mnras, 482, 5138

\bibitem[Bekki(2010)]{bekki10}
Bekki, K.\ 2010, \apjl, 723, L99

\bibitem[Briley et al.(1989)]{briley89}
Briley, M.~M., Smith, G.~H., Bell, R.~A., Oke, J.~B., \& Hesser, J.~E.\ 1992, \apj, 387, 612

\bibitem[Brown et al.(2018)]{gaiadr2}
Brown, A.G.A., Vallenari, A., Prusti, T.\ et al.\ 2018, \aap, 616, A1

\bibitem[Campbell et al.(2013)]{campbell13}
Campbell, S.~W., D'Orazi, V., Yong, D.\ et al. 2013, \nat, 498, 198

\bibitem[Carretta(2014)]{carretta14} 
Carretta, E.\ 2014, \apjl, 795, L28

\bibitem[Carretta et al.(2009)]{carretta09} 
Carretta E.,  Bragaglia, A., Gratton, R.G., Lucatello S., Cantanzaro G. et al. 2009, \aap, 505, 117

\bibitem[Cassisi \& Salaris(2013)]{cassisi13}
Cassisi, S., \& Salaris, M.\ 2013, Old Stellar Populations: how to study
the fossil record of galaxy formation (Berlin:Wiley-VCH)

\bibitem[Castelli(2005)]{castelli05}
Castelli, F.\ 2005, Mem.\ S.A.It.\ Suppl.\ 8, 25

\bibitem[Cohen et al.(2002)]{cohen02}
Cohen, J.~G., Briley, M.~M., Stetson, P.~B.\ 2002, \aj, 123, 2525

\bibitem[Crestani et al.(2019)]{crestani19}
Crestani, J., Alves-Brito, A., Bono, G. et al.\ 2019, \mnras, 487, 5463

\bibitem[Dotter et al.(2008)]{dotter08}
Dotter, A., Chaboyer, B., Jevremovi{\'c}, D.\ et al.\ 2008, \apjs, 178, 89

\bibitem[Dupree \& Avertt(2013)]{dupree13}
Dupree, A.~K., \& Avrett, E.~H.\ 2013, \apjl, 773, L28

\bibitem[Gratton et al.(2015)]{gratton15}
Gratton, G.~R., Lucatello, S., Sollima A.\ et al.\ 2015, \aap, 573, A92

\bibitem[Grundahl et al.(1999)]{grundahl99} 
Grundahl, F., Catelan, M., Landsman, W.~B., Stetson, P.~B., \& Andersen, M.~I.\ 1999, \apj, 524, 242 

\bibitem[Harris(1996)]{harris96}
Harris, W.\ E.\ 1996, \aj, 112, 1487

\bibitem[Ivans et al.(2001)]{ivans01}
Ivans, I.~I., Kraft, R.~P., Sneden, C.\ et al.\ 2001, \aj, 122, 1438

\bibitem[Kurucz(1970)]{kurucz70}
Kurucz, R., L.\ 1970, SAOSR, 309, 291

\bibitem[Kurucz(2005)]{kurucz05}
Kurucz, R., L.\ 2005, Mem.\ S.A.It.\ Suppl.\ 8, 14

\bibitem[Lagioia et al.(2018)]{largioia18}
Lagioia, E.~P., Milone, A.~P., Marino, A.~F., et al.\ 2018, \mnras, 475, 4088

\bibitem[Lee(2010)]{lee10} Lee, J.-W.\ 2010, \mnras, 405, L36

\bibitem[Lee(2015)]{lee15} Lee, J.-W.\ 2015, \apjs, 219, 7

\bibitem[Lee(2016)]{lee16} Lee, J.-W.\ 2016, \apjs, 226, 16

\bibitem[Lee(2017)]{lee17} Lee, J.-W.\ 2017, \apj, 844, 77

\bibitem[Lee(2018)]{lee18} Lee, J.-W.\ 2018, \apjs, 238, 24

\bibitem[Lee(2019a)]{lee19a} Lee, J.-W.\ 2019a, \apj, 872, 41

\bibitem[Lee(2019b)]{lee19b} Lee, J.-W.\ 2019b, \apjl, 875, 27

\bibitem[Lee \& Carney(1999)]{lc99} 
Lee, J.-W., \& Carney, B.~W.\ 1999, \aj, 117, 2868 

\bibitem[Lee et al.(2009a)]{jwlnat}
Lee, J.-W., Kang, Y.-W., Lee, J., \& Lee, Y.-W.\ 2009a, \nat, 462, 480

\bibitem[Lee et al.(2009b)]{lee09}
Lee, J.-W., Lee, J., Kang, Y.-W., et al.\ 2009b, \apjl, 695, L78

\bibitem[Lee et al.(2014)]{lee14}
Lee, J.-W., L\'opez-Morales, M., Hong, K.\ et al.\ 2014, \apjs, 210, 6

\bibitem[Lee \& Pogge(2016)]{lp16}
Lee, J.-W., \& Pogge, R.\ 2016, JKAS, 49, 289

\bibitem[S.-G.\ Lee(1999)]{sglee99} Lee, S.-G.\ 1999, \aj, 118, 920

\bibitem[Lim et al.(2016)]{lim16}
Lim, D., Lee, Y.-W., Pasquato, M.\ et al.\ 2016, \apj, 832, 99

\bibitem[Maas et al.(2019)]{maas19}
Maas, Z.~G., Gerber, J.~M., Deibel, A.\ \& Pilachowski, C.~A.\ 2019, \apj, 878, 43

\bibitem[Marino et al.(2018)]{marino18}
Marino, A.~F., Milone, A.~P., Casagrande, L., et al.\ 2018, \apjl, 863, L33

\bibitem[Marino et al.(2017)]{marino17}
Marino, A.~F., Milone, A.~P., Yong, D., et al.\ 2017, \apj, 843, 66

\bibitem[Marino et al.(2008)]{marino08}
Marino, A.~F., Villanova, S., Piotto, G.\ et al.\ 2008, \aas, 490, 625

\bibitem[Milone et al.(2018)]{milone18}
Milone, A.~P., Marino., A.~F., Mastrobuono-Battisti, A., \& Lagioia, E.~P.\ 2018, \mnras, 479, 5005

\bibitem[Milone et al.(2015)]{milone15}
Milone, A.~P., Marino, A.~F., Piotto, G.\ et al.\ 2015, \apj, 808, 51

\bibitem[Milone et al.(2017)]{milone17}
Milone, A.~P., Piotto, G., Renzini, A.\ et al.\ 2017, \mnras, 464, 3636

\bibitem[Norris et al.(1981)]{norris81}
Norris, J., Cottrell, P.~L., Freeman, K.~C., \& Da Costa, G.~S.\ 1981, \apj, 244, 205

\bibitem[Piotto et al.(2015)]{piotto15}
Piotto. G., Milone, A.~P., Bedin, L.~R.\ et al.\ 2015, \aj, 149, 91

\bibitem[Sbordone et al.(2004)]{sbordone04}
Sbordone, L., Bonifacio, P., Castelli, F., \& Kurucz, R.~L.\ 2004, Mem.\ S.A.It.\ Suppl.\ 5, 93

\bibitem[Smith(1987)]{smith87}
Smith, G.~H.\ 1987, \pasp, 99, 67

\bibitem[Smolinski et al.(2011)]{smolinski11}
Smolinski, J.~P., Martell, S., Beers, T.~C., et al.\ 2011, \apj, 142, 126

\bibitem[Sneden(1973)]{sneden73} 
Sneden, C.\ 1973, \apj, 184, 839

\bibitem[Sneden, Ivans, \& Kraft(2000)]{sneden00}
Sneden, C., Ivans, I.~I., \& Kraft, R.~P.\ 2000, \memsai, 71, 657

\bibitem[Stetson(1987)]{pbs87} 
Stetson P.\ B.\ 1987, \pasp, 99, 191

\bibitem[Stetson(1994)]{pbs94} 
Stetson P.\ B.\ 1994, \pasp, 106, 250

\bibitem[Stetson \& Harris(1988)]{sh88} 
Stetson P.\ B., \& Harris, W.\ E.\ 1988, \aj, 96, 909

\bibitem[Zachairias et al.(2004)]{nomad}
Zacharias, N., Monet, D.\ G., Levine, S.\ E., et al.\ 2004, AAS, 205, 4815

\end{thebibliography}
\end{document}